\documentclass[aps,prl,twocolumn,groupedaddress,showpacs]{revtex4-1}
\usepackage{amsmath,amssymb}
\usepackage{graphicx}
\usepackage{natbib}
\usepackage{bm}

\begin{document}
\title{Spectroscopy for cold atom gases in periodically phase-modulated
optical lattices}

\author{Akiyuki Tokuno}
\affiliation{DPMC-MaNEP, University of Geneva, 24 Quai Ernest-Ansermet
CH-1211 Geneva, Switzerland.}

\author{Thierry Giamarchi}
\affiliation{DPMC-MaNEP, University of Geneva, 24 Quai Ernest-Ansermet
CH-1211 Geneva, Switzerland.}

\date{\today}

\begin{abstract}
 The response of cold atom gases to small periodic phase modulation of
 an optical lattice is discussed.
 For bosonic gases, the energy absorption rate is given, within linear
 response theory, by the imaginary part of the current autocorrelation
 function.
 For fermionic gases in a strong lattice potential, the same correlation
 function can be probed via the production rate of double occupancy.
 The phase modulation gives thus direct access to the conductivity of
 the system, as a function of the modulation frequency. We give an
 example of application in the case of bosonic systems at zero
 temperature and discuss the link between the phase- and amplitude-
 modulation.
\end{abstract}

\pacs{67.85.-d,05.30.Jp,03.75.Ss,05.30.Fk,71.10.Fd,78.47.-p}

\maketitle

Cold atomic systems have proven to be remarkable laboratories to study
several effects of strongly correlated systems.
In particular the control of parameters, kinetic energy in an optical
lattice and interaction using a Feshbach resonance, allows us to
potentially use them as quantum simulators, with considerable success
both for pure and disordered
systems~\cite{Bloch/RevModPhys.80.2008:review,Sanchez-Palencia/NatPhys.6.2010}.
However, in addition to realizing the systems, the ability to probe it
is important.
Because of the electrical charge neutrality of cold atoms, unlike electron
systems, they are insensitive to the usual electromagnetic probes.
This makes it potentially difficult to probe correlations in such
systems.
To overcome this issue, several probes have been proposed
besides the standard time of flight (TOF) experiment such as Bragg
spectroscopy~\cite{Stenger/PRL.82.1999,Rey/PRA.72.2005,Hagley/PRL.83.1999,Pupillo/PRA.74.2006}
to measure the dynamic structure factor, radio frequency spectroscopy
measurement~\cite{Regal/Nat.424.2003,Stoeferle/PRL.96.2006} to count the
number of molecules formed by the Feshbach resonance, shot noise
measurement~\cite{Altman/PRA.70.2004,Greiner/PRL.94.2005,Foelling/Nat.434.2005}
for the density-density correlation function or momentum-resolved Raman
spectroscopy~\cite{Dao/PRA.80.2009,Stewart/Nat.454.2008} for the
single-body spectrum function.

Among the various spectroscopic probes a particularly simple probe
consists in changing periodically the amplitude of the optical
lattice~\cite{Stoeferle/PRL.92.2004,Schori/PRL.93.2004}.
The energy absorbed by such a modulation can be estimated from the TOF
image.
The corresponding theory of the energy absorption rate (EAR)
spectrum~\cite{Iucci/PRA.73.2006,Kollath/PRL.97.2006}, was shown to give
access both to the Mott-insulating (MI) gap and to the kinetic energy
correlations in the system.
Although measuring the EAR by the TOF was possible for bosons, a similar
measure was highly inconvenient for fermions.
It was proposed~\cite{Kollath/PRA.74.2006} that a measurement of the
doublon production rate (DPR) in response to the amplitude modulation
would give access to the same information.
Such a measure was successfully implemented for fermionic
systems~\cite{Joerdens/Nat.455.2008,Strohmaier/PRL.104.2010,Esslinger/AnnuRevCondMattPhys.1.2010:review,Sensarma/PRL.103.2009}.
The amplitude modulation of the optical lattice coupled either to EAR or
to DPR is thus a simple but powerful and versatile probe.

In this Letter we propose an alternative probe, based on a phase
modulation of an optical lattice potential.
Such a modulation is known to lead to a
current~\cite{Drese/ChemPhys.217.1997,Arlinghaus/PRA.81.2010,Graham/PRA.45.1992,Citro/PRA.83.2011}
or to band narrowing~\cite{Madison/PRL.81.1998,Lignier/PRL.99.2007}.
Here we use the phase modulation in connection with EAR or DPR techniques,
to analyze the spectrum of the system.
We show that such a probe gives access to the current autocorrelation
function and is thus analogous to optical conductivity
measurements in condensed matter systems, allowing a very close
comparison at the experimental level between the two domains.
We illustrate the use of such a probe by some examples for bosonic
gases and compare with the spectrum obtained by the amplitude modulation
spectroscopy.

\begin{figure}[bp]
  \begin{center}
   \includegraphics[scale=0.70]{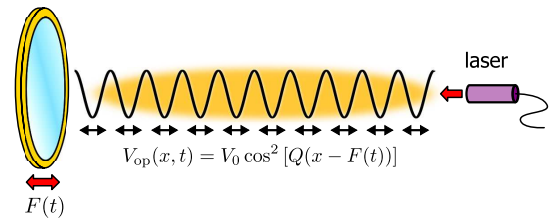}
   \caption{(color online). A schematic showing the setup of the
   periodic phase modulation of an optical lattice. 
   The incident laser and the reflected one forms the standing wave
   which corresponds to an optical lattice.
   The lattice potential follows mirror oscillation, and consequently
   the phase is modulated.
   }
   \label{fig:scheme}
  \end{center}
\end{figure}

Let us first describe our proposed probe: The optical lattice potential
is created by shining laser against a mirror.
If the mirror is stationary, the created $D$-dimensional optical lattice
is given as
$V_{\rm op}(\bm{r})=V_0\sum_{\mu=1}^{D}\cos^2(Q_{\mu}r_{\mu})$ where
$\bm{Q}=(Q_1,\cdots,Q_D)$ is a wave vector of the optical lattice.
One can modulate the phase of an optical lattice potential by
oscillating the mirror as shown in Fig.~\ref{fig:scheme}.
The lattice potential in the laboratory frame is modified as
$V_{\rm op}(\bm{r},t)=\sum_{\mu=1}^{D}V_0\cos^2[Q_{\mu}(r_{\mu}-F_{\mu}(t))]$
where ${\bm F}(t)$ represents the oscillation of the phase.
It is convenient to switch to the comoving frame by the gauge transform
$U(t)=\exp(iM\bm{F}(t)\cdot\bm{J}/\hbar)$ 
where $M$ is a mass of the atoms and $\bm{J}$ the current
operator~\footnote{For simplicity, we considered single-species atom
systems. But, the formalism can be extended to mixtures. Then the
corresponding gauge transform is given by
$M\bm{J}\rightarrow\sum_{\alpha}M_\alpha J_{\alpha}$
where $\alpha$ denotes the species of atoms.}.
In the comoving frame, the lattice becomes a stationary one,
$V_{\rm op}(\bm{r})$, and an additional term, which reflects the
inertial force, emerges in the Hamiltonian:
\begin{equation}
  H(t)=H_0-M\dot{\bm{F}}(t)\cdot \bm{J},
  \label{eq:Hamiltonian}
\end{equation}
where $H_0$ is the Hamiltonian of the interacting system in the optical
lattice $V_{\rm op}(\bm{r})$.
Carrying out the further gauge transform
$U'(t)=\exp[iNM\int^{t}\!\! dt' \dot{\bm F}^2(t')/2\hbar]$
where $N$ is the total atom number, we can find an expression identical
to that of a {\it charged} particle in an electromagnetic field,
the atomic mass corresponding to the charge.
The vector potential is given as $\bm{A}_{\rm ext}=\dot{\bm F}(t)$ with
$\bm{\nabla}\cdot\bm{A}_{\rm ext}=0$ and the scalar potential is zero
$\phi_{\rm ext}=0$~\cite{Drese/ChemPhys.217.1997}.
The system thus behaves as {\it charged} particles under an electric
field $\bm{E}_{\rm ext}(t)=-\ddot{\bm{F}}(t)$.
Hereafter we set $\bm{F}(t)=\bm{f}\cos(\omega t)$.

Let us first consider bosonic atom cases.
One can then measure the EAR by similar techniques than for the
amplitude modulation~\cite{Stoeferle/PRL.92.2004}.
The EAR is given by  the time average of the absorbed energy:
$R(\omega)=\frac{\omega}{2\pi}\int_{T}^{T+2\pi/\omega}\!\!dt\langle\dot{H}(t)\rangle$
where $\langle\cdots\rangle$ denotes the statistical average with the
Hamiltonian~(\ref{eq:Hamiltonian}).
The EAR within the linear response theory is given by
\begin{equation}
  R_{\rm PM}(\omega)
   =-\frac{M^2\omega^3}{2\hbar}
     \sum_{\mu,\nu=1}^{D}
       f_\mu f_\nu
       \Im\tilde{\Pi}_{\mu\nu}^{\rm R}(\omega),
  \label{eq:PMEAR}
\end{equation}
where $\tilde{\Pi}_{\mu\nu}^{\rm R}(\omega)$ is the Fourier transform of
the retarded current correlation function
for the Hamiltonian $H_0$.
Note that for the EAR, due to
$\dot{H}(t)=M\omega^2\cos(\omega t)\bm{f}\cdot\bm{J}$,
one can automatically derive the second-order response of $R(\omega)$ in
terms of $f_\mu$ within the first order perturbation theory.
Note, as shown in Eq.~(\ref{eq:Hamiltonian}), that in order to stay
within linear response a small modulation is necessary.
In particular one needs $|F_{\mu}(t)|\ll Q_{\mu}^{-1}$, thus
a modulation amplitude smaller than a lattice constant.
This is something difficult but achievable with the
current experimental technique~\cite{Reischl/PRA.72.2005}.
For the higher frequency the smaller amplitude
needed to stay within the linear response and away from the
dynamically induced phase
transition~\cite{Eckardt/PRL.95.2005,Eckardt/EuroPhysLett.80.2007}.
We will confine our analysis in the following to such a regime for which
the EAR gives direct access to $\tilde{\Pi}_{\mu\nu}^{\rm R}(\omega)$.

Since the EAR directly gives $\tilde{\Pi}_{\mu\nu}^{\rm R}(\omega)$, it
is immediately related to the ``optical
conductivity''~\cite{Mahan/2000:book}.
While the zero frequency part of the Drude peak in the conductivity
will be suppressed in the EAR due to the factor $\omega^3$ in
Eq.~(\ref{eq:PMEAR}), all the other features, at finite frequency,
are perfectly reproduced.
It is thus a particularly useful quantity to make comparison with
similar phenomena in condensed matter systems, or to probe the
physics of disordered systems, for which transport is the prime
probe.

It is interesting to compare this result to another one obtained for the
amplitude modulation~\cite{Iucci/PRA.73.2006}, in the same linear response regime.
In the later case the EAR is either given by the density correlation
function for weak optical lattices or by the kinetic energy one for
strong optical lattices~\cite{Iucci/PRA.73.2006,Sensarma/PRL.103.2009}.
The different representation of perturbation operator comes from the
fact that the energy scale of the amplitude modulation also goes beyond the
chemical potential as the energy scale of the lattice potential
$V_{\rm op}(\bm{r})$ increases.
On the contrary, in the case of the phase modulation, the perturbation in
Eq.~(\ref{eq:Hamiltonian}) does not follow the energy scale of the
lattice potential.
Therefore, Eq.~(\ref{eq:PMEAR}) is \emph{independent} of the strength
of the lattice potential.
This is a definite advantage of the phase modulation, which is always
related to the
same physical quantity irrespectively of the strength of the optical
lattice.
Another important difference between the phase and amplitude
modulation comes from their symmetries.
Indeed, for example, in the case of a 1D strong lattice the amplitude
modulation would
correspond to the kinetic energy correlation function which is given by
$T\propto\sum_k\cos(k) b^\dagger_k b_k$ while the current is given
by $J\propto\sum_k\sin(k) b^\dagger_k b_k$.
For the parity inversion, $\bm{J}\rightarrow -\bm{J}$ while $T$ is
invariant.
This affects the selection rules.
The phase modulation perturbation probes the transitions from a state to
an opposite parity state.
In contrast, the parity is preserved in the transition due to the
amplitude modulation.
Thus, both modulations complement each other in the way they probe
excited states, and lead in general to different results.

For fermions, as for the amplitude modulation, the EAR is not a
convenient way to probe the consequences of the modulation.
We follow here the same approach as in~\cite{Kollath/PRA.74.2006} and
show that for the phase modulation the measurement of the DPR gives
essentially the same information as the EAR.
We assume that $H_0$ in Eq.~(\ref{eq:Hamiltonian}) is described by the
Hubbard model, $H_0=-t_{\rm H}\sum_{\sigma,\langle{i,j}\rangle}c^{\dagger}_{i\sigma}c_{j\sigma}+U\sum_{j}n_{j\uparrow}n_{j\downarrow}$.
The number of doubly occupied sites is defined as
$N_{\rm D}(t)=\langle{V}\rangle/U$,
and can be rewritten as
$N_{\rm D}(t)=\frac{1}{U}[\langle{H(t)}\rangle-\langle{T}\rangle+M\dot{\bm{F}}\cdot\langle\bm{J}\rangle]$,
where $T$ and $V$ are the kinetic energy
and interaction terms in the above Hamiltonian.
The production rate is defined as the time average of
$\dot{N}_{\rm D}(t)$ for a period:
$P(\omega)=\frac{\omega}{2\pi}\int_{t}^{t+2\pi/\omega}\!\!dt'\dot{N}_{\rm D}(t')$.
A second-order perturbation expansion in $\bm{F}(t)$ gives for
the productions rate
\begin{equation}
P_{\rm PM}(\omega)=R_{\rm PM}(\omega)/U,
\label{eq:prodd}
\end{equation}
which shows the direct relation between the EAR and the DPR for the
phase modulation. This shows that DPR gives also access to the optical
conductivity for these system.

The results (\ref{eq:PMEAR}) and the equivalence of the DPR to the EAR
(\ref{eq:prodd}) are thus our main proposal to use the phase modulation
of the optical lattice to measure the optical conductivity of
interacting systems in a cold atom context.

Let us now examine an example of the phase modulation technique.
For the sake of simplicity we take a repulsively interacting 1D bosonic
atom system at zero temperature.
The unperturbed Hamiltonian in Eq.~(\ref{eq:Hamiltonian}) is
written as
\begin{equation}
  H_0
  =\int\!\! dx
   \left[
    \psi^\dagger
    \left(
      -\frac{\hbar^2}{2M}\partial_x^2
      -\mu
      +V_{\rm op}(x)
    \right)
    \psi
    +\frac{g}{2}\rho^2
   \right],
  \label{eq:unperturbedHamiltonian}
\end{equation}
where $V_{\rm op}(x)=V_0\cos^2 Qx$ is a 1D lattice potential, and $g$
an interaction parameter.
The field $\psi(x)$ and $\rho(x)$ are, respectively, the annihilation
and density operators.

For a shallow lattice potential, $V_0\ll \mu$, $V_{\rm op}(x)$ can be
regarded as a perturbation, and then the Hamiltonian
(\ref{eq:unperturbedHamiltonian}) can be rewritten via the
bosonization~\cite{Haldane/PRL.47.1981,Giamarchi/2004:book},
$\psi^{\dagger}(x)\sim\sqrt{\bar{\rho}}e^{-i\theta(x)}$
and $\rho(x)\sim\bar{\rho}-\partial_x\varphi(x)/\pi+2\bar{\rho}\cos[2\pi\bar{\rho}x-2\varphi(x)]$,
where $\bar{\rho}$ is the mean density of the system.
Retaining only the most relevant term generated by the presence of
$V_{\rm op}(x)$, one can obtain
\begin{equation}
  H_{\rm eff}
  =H_{\rm TL}
   +\lambda\int\!\! dx\cos \left[2(Q-\pi\bar{\rho})x+2\varphi(x)\right],
  \label{eq:sinegordon}
\end{equation}
where $H_{\rm TL}=\frac{\hbar v}{2\pi}\int\!\!dx\left[K(\partial_x\theta(x))^2+K^{-1}(\partial_x\varphi(x))^2\right]$
is the Tomonaga-Luttinger (TL) Hamiltonian.
The phase fields $\theta(x)$ and $\varphi(x)$ represent, respectively,
the phase and density fluctuations of bosons.
For an arbitrary repulsion $K$ runs from $\infty$ to unity as the
interaction increases, and $K=1$ and $\infty$ correspond to the Tonks
gas and the noninteracting bosons, respectively.
Thus in the boson systems the low-energy physics is mainly governed by
the cosine term in Eq.~(\ref{eq:sinegordon}).
Furthermore, the current is written as
$J=vK\int\!\!dx\ \Pi(x)$ where $\Pi(x)$ is the canonically
conjugate momentum to $\varphi(x)$.
In the incommensurate case, i.e., $\pi\bar{\rho}\ne Q$, the
cosine term in Eq.~(\ref{eq:sinegordon}) vanishes, and the effective
theory is the TL liquid.
For $\pi\bar{\rho}= Q$, the model (\ref{eq:sinegordon}) becomes a
sine-Gordon model for which the cosine term is relevant for $K<2$,
leading to a MI gap in the excitation spectrum~\cite{Giamarchi/2004:book}.
We will consider this last case in what follows.

The conductivity can be calculated using the methods in
Refs.~\cite{Giamarchi/2004:book,Giamarchi/PRB.46.1992}.
To determine the large frequency behavior of the 1D current correlation
function $\tilde{\Pi}^{\rm R}_J(\omega)$, we use the memory function
method, which gives correctly $\tilde{\Pi}^{\rm R}_J(\omega)$ for large
frequency compared to the MI gap.
The memory function
$M(\omega)\equiv\omega\tilde{\Pi}^{\rm R}_{J}(\omega)/[\tilde{\Pi}^{\rm R}_{J}(0)-\tilde{\Pi}^{\rm R}_{J}(\omega)]$
can be approximated as
$M(\omega)\approx [\tilde{\Pi}^{\rm R}_F(0)-\tilde{\Pi}^{\rm R}_F(\omega)]/\omega\tilde{\Pi}^{\rm R}_J(0)$
where $\tilde{\Pi}^{\rm R}_F(\omega)$ is the retarded correlation
function of $F(t)=[H_0,J(t)]$.
From the Hamiltonian~(\ref{eq:sinegordon}), $F$ is given
by $F=i2vK\lambda\int\!\!dx\sin[2(Q-\pi\bar{\rho})+2\varphi(x)]$,
and $M(\omega)\sim\omega^{2K-3}$ is immediately computed.
In the gapless ($K>2$) case, negligible $M(\omega)/\omega$ for small
$\omega$ leads to $\tilde{\Pi}^{\rm R}_{J}(\omega)\propto\omega^{2K-5}$.
As a result, it is found that the EAR spectrum for small $\omega$ and
$K>2$ behaves as $R_{\rm PM}(\omega) \propto \omega^{2K-2}$.
This is to be compared with the amplitude modulation
result~\cite{Iucci/PRA.73.2006}
$R_{\rm AM}(\omega) \propto \omega^{2K-1}$ for weak lattices.
A similar result is obtained in the large-$\omega$ limit for the massive
case $K<2$.
In the gapful case the cosine is relevant, and the conductivity, i.e.,
the phase modulation response will be zero below the
gap~\cite{Giamarchi/2004:book}.
This example thus shows differences between the phase and amplitude
modulation. This difference has two origins; one is the trivial different prefactors
of the correlation functions ($\omega^3$ for the phase modulation and
$\omega$ for the amplitude one).
More importantly and as discussed above, the main difference comes from
the perturbation coupling to two different operators: namely the current
for the phase modulation and the density for the amplitude modulation in
the shallow lattice limit.

\begin{figure}[t]
  \begin{center}
   \includegraphics[scale=0.9]{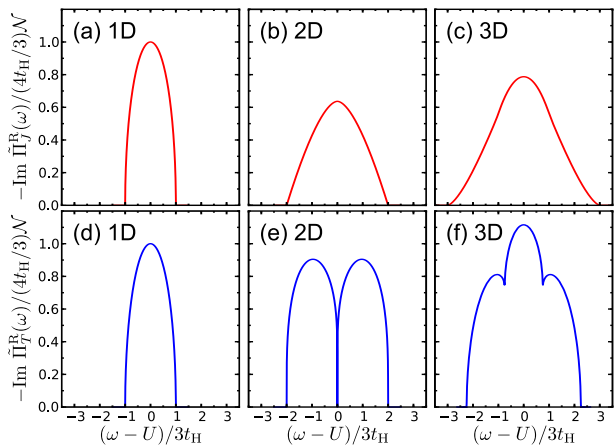}
   \caption{(color online). The imaginary part of the current [(a)-(c)],
   and the kinetic-energy [(d)-(f)] correlation functions in the bosonic
   Mott insulator for $t_{\rm H}/U=0.01$ at zero temperature.
   The lattice constant and $\hbar$ have been taken to be unity.
   For 1D [(a) and (d)], these two correlation functions are identical
   while they are qualitatively different for 2D [(b) and (e)] and for
   3D [(c) and (f)].
   }
   \label{fig:spectrum}
  \end{center}
\end{figure}

We now consider a strong lattice potential.
Then, the system is well described by a lattice model: $H_0$
is given by the Bose-Hubbard Hamiltonian $H_{\rm eff}=T+V$
where $T=-t_{\rm H}\sum_j [b_{j+1}^{\dagger}b_{j}+{\rm h.c.}]$
and $V=\frac{U}{2}\sum_j n_j \left(n_j-1 \right)$.
$b_j^{\dagger}$ and $n_j$ are, respectively, the creation and
number operators for a bosonic atom at the
$j$th site, and $t_{\rm H}$ and $U$ are the hopping parameter and
on-site interaction, respectively.
For an incommensurate filling, the ground state is in the gapless
superfluid (SF) phase.
For filling of one particle per site, the SF-MI transition occurs
at $U/t_{\rm H}=1.92$~\cite{Kuhner/PRB.58.1998}.
In the SF phase, the low-energy physics is governed by the TL
liquid~\cite{Giamarchi/2004:book}.
In the MI phase, an energy gap opens, and the low-energy physics is no
longer described by the TL liquid.
The lowest energy excitation above the gap is formed by a pair of atoms
at the same site (doublon) and an empty site (holon).
In the limit $t_{\rm H}/U\rightarrow 0$, the pair excitations are
${\cal N}({\cal N}-1)$-fold degenerate where ${\cal N}$ is the number
of lattice sites.
For finite but small $t_{\rm H}/U$, the degenerate energy levels split,
and an energy band whose width is about $t_{\rm H}$ is formed.
This band leads in the phase and amplitude modulation spectrums to a
peak around $\omega\approx U/\hbar$ as shown in
Fig.~\ref{fig:spectrum}. 

Using degenerate perturbation theory \cite{Iucci/PRA.73.2006}, the
EAR spectrum can be calculated.
Let $|d_Rh_r\rangle$ be a pair state of the doublon and holon at $R$th
and $r$th site, respectively, which is an exact eigenstate of $V$.
We represent $T$ in the Hilbert space spanned by $|d_Rh_r\rangle$.
The eigenstate of $T$ is 
$|{l,l'}\rangle= \frac{\sqrt{2}}{\cal N}\sum_{R=1}^{\cal N}\sum_{r=1}^{{\cal N}-1}e^{i(p_lR+\arg[w_{p_l}] r)}\sin(q_{l'}r)|d_{R}h_{R+r}\rangle$
where $w_{p_l}=1+2e^{ip_l}$, $p_l=2\pi l/{\cal N}$ and
$q_{l'}=\pi l'/{\cal N}$ ($l=1,\cdots,{\cal N}$ and
$l'=1,\cdots,{\cal N}-1$).
The corresponding eigenenergy is 
$E_{l,l'}=U-t_{\rm H}|w_{p_l}|\cos{q_{l'}}$.
The retarded correlation function of ${\cal O}$ for
$\omega>0$ is expressed as
$\Im\tilde{\Pi}^R_{\cal O}(\omega)=-\pi\hbar\sum_{n}\left|\langle n|{\cal O}|0\rangle\right|^2\delta(\hbar\omega-E_{n})$
where $|n\rangle$ and $|0\rangle=\prod_{j}b^{\dagger}_{j}|vac\rangle$,
respectively, denotes an intermediate state and the MI ground state.
Restricting the intermediate states onto $|l,l'\rangle$, both the
amplitude modulation and the phase modulation can be computed.
In the ${\cal N}\rightarrow \infty$ limit they turn out to be identical
and can be written as
$\Im\tilde{\Pi}^{\rm R}_{J}(\omega)=-{\cal N}\frac{\pi^2t_{\rm H}}{\hbar Q^2}\pi(\omega)$
and $\Im\tilde{\Pi}^{\rm R}_{T}(\omega)=-{\cal N}\hbar t_{\rm H}\pi(\omega)$~\cite{Iucci/PRA.73.2006}
where
$\pi(\omega)=\frac{4}{3}\sqrt{1-[(\hbar\omega-U)/3t_{\rm H}]^2}$,
as shown in Fig.~\ref{fig:spectrum}.
Therefore, the appropriately scaled EAR, i.e.,
$\omega^{-2}R_{\rm PM}(\omega)$, is identical to $R_{\rm AM}(\omega)$. 
In this particular case, $\tilde{\Pi}^{\rm R}_{J}(\omega)$ and
$\tilde{\Pi}^{\rm R}_{T}(\omega)$ are identical.
This is a peculiar feature of the 1D MI excitation spectrum, linked to
the fact that in 1D the hole and doublon
cannot cross each other during their motion.
Thus qualitative difference must appear in the 2D and 3D cases.
We thus compute $\tilde{\Pi}^{\rm R}_{J}(\omega)$ and
$\tilde{\Pi}^{\rm R}_{T}(\omega)$ for 2D and for 3D by using an
diagrammatic approach. We consider a doublon and holon with an infinite repulsive interaction 
which implements the constraint that the two particles cannot be at the same point except when they recombine. 
For 1D this method is in full agreement with~\cite{Iucci/PRA.73.2006,Iucci/unpublished}. 
The result is also shown in Fig.~\ref{fig:spectrum}.
One clearly sees the difference between the two modulations. Note that the anomalous structures in the amplitude 
modulations are related to the Van Hove singularities in the density of states. 

In summary we have proposed in this Letter to use small periodic phase
modulation of an optical lattice to probe for the current autocorrelation function.
The consequences of the modulation can be measured either by probing the
absorbed energy of the system (for bosons) or by measuring the production
rate of doubly occupied sites (for fermions).
Such a phase modulation probe gives direct access to the frequency
dependent conductivity of the system.

\begin{acknowledgments}
 The authors are grateful to M. A. Cazalilla, A. Iucci, M. Oshikawa
 and C. Salomon for useful discussions.
 This work was supported by the Swiss National Foundation under MaNEP
 and division II.
\end{acknowledgments}

\bibliographystyle{apsrev4-1}
%

\end{document}